# Blockchain-based Medical Image Sharing and Automated Critical-results Notification: A Novel Framework

Jiyoun Randolph*, Md Jobair Hossain Faruk†, Hossain Shahriar*, Maria Valero*
Liang Zhao*, Nazmus Sakib*, Bilash Saha*

*Department of Information Technology, Kennesaw State University, GA, USA
†Department of Software Engineering and Game Development, Kennesaw State University, GA, USA
{jrando18, mhossa21, bsaha}@students.kennesaw.edu | {hshahria, mvalero2, lzhao10, nsakib1}@kennesaw.edu

*Abstract*— In teleradiology, medical images are transmitted to offsite radiologist for interpretation and the dictation report is sent back to the original site to aid timely diagnosis and proper patient care. Although teleradiology offers great benefits including time and cost efficiency, after-hour coverages, and staffing shortage management, there are some technical and operational limitations to overcome in reaching its full potential. We analyzed the current teleradiology workflow to identify inefficiencies. Image unavailability and delayed critical result communication stemmed from lack of system integration between teleradiology practice and healthcare institutions are among the most substantial factors causing prolonged turnaround time. In this paper, we propose a blockchain-based medical image sharing and automated critical-results notification platform to address the current limitation. We believe the proposed platform will enhance efficiency in workflow by eliminating the need for intermediaries and will benefit patients by eliminating the need for storing medical images in hard copies. While considerable progress was achieved, further research on governance and HIPAA compliance is required to optimize the adoption of the new application. Towards an idea to a working paradigm, we will implement the prototype during the next phase of our study.

## I. INTRODUCTION

Teleradiology is a branch of telemedicine [1]. It refers to a practice where radiologists interpret radiological studies offsite from where the studies are generated. In teleradiology, medical images are transmitted electronically to offsite radiologists (teleradiologists) and dictation reports are sent back to the original site to aid proper patient care [2]. Teleradiology offers great benefits of time and cost efficiency, after-hours coverages, and staffing shortage management to improve the quality of patient care, efficiency of the healthcare system, and productivity of radiologists [3]. It is widely used for hospitals, imaging centers, emergency facilities, and mobile x-ray services and commercial radiology outsourcing enterprises have been emerging. The field's growth has accelerated with massive growth in telehealth, especially during the recent COVID-19 pandemic. With this trend, the global teleradiology market size was estimated to be $3.453 billion in 2019 and is expected to reach $8.024 billion by 2026 [4].

Despite their apparent promise, there are some limitations in reaching the full potential of teleradiology. Systems integration has been one of the main challenges as image management system resides in the onsite information system security domain and teleradiology practice has its own system [5]. Working with disparate systems causes challenges in medical image transmission and critical result communication between teleradiology practice and other health care institutions. In addition, teleradiologists often do not have direct access to the onsite Radiology Information System (RIS) or Picture Archiving and Communication System (PACS), instead, use web-based image distribution systems [3].

This can be problematic not only because the web-based image distribution systems offer limited functions compared to those used by onsite radiologists but also teleradiologists can only provide provisional reports in some cases due to their limited access to patient history and prior exams for comparison. Furthermore, current image transmission operates on the "demand-push" model, where the sending site needs to initiate image transmissions. Hence, teleradiologists do not have direct access to additional data such as prior reports if those reports have not been sent by the sending site and must request additional images [3]. Lastly, lack of effective means of communication between interpreting radiologists and referring physicians/care teams can affect patient outcomes significantly with critical conditions where timely treatment can drastically improve survival rates and reduce complications.

Identified challenges not only cause inefficiencies in the workflow but also are interruptions in radiologists' workflow. This paper will examine the current teleradiology workflow and identify inefficiencies in image sharing and critical-results communication. We propose a blockchain-based model for image sharing and critical-results communication to address current limitations in teleradiology.

The remainder of the paper is structured as follows: Section II provides an overview of previous research on healthcare blockchain applications and a brief introduction to Hyperledger Fabric followed by discusses challenges in the current teleradiology workflow in Section III. Section IV presents a proposed architecture for a blockchain-based medical image sharing and automated critical-results notification application while Section V discusses the advantages and limitations of the proposed application, and Section VI concludes the paper.

## II. RELATED WORK

*A. Literature Review*

Although there is an increasing interest in blockchain technology from healthcare, there are limited use cases and research on the use of blockchain in healthcare compared to other industries, especially in radiology. Healthcare blockchain applications are primarily at the conceptual stage, and many of them focus on managing EHR (Electronic Health Records) or medical research data. Scores of research work have been conducted in healthcare by adopting blockchain technology that includes supply chain management in healthcare and patient-driven interoperability [6], [7].

A. Azaria et el. [8] proposed blockchain-based medical data access and permission management system, MedRec. MedRec is the first to develop a fully functional blockchain prototype to address the challenges in current EHR systems such as data fragmentation, data accessibility, system interoperability, patient agency, and quality and quantity of medical research data. Each node in the blockchain network only stores copies of authorization data while medical records are stored on external databases. It utilizes data pointers to retrieve distributed medical records and Ethereum's smart contracts for permissions, data sharing, and mining. Its smart contract structure consists of Register Contract (RC), Patient-Provider Relationship Contract (PPR), and Summary Contract (SC). RC regulates new identity registration and maps IDs with their Ethereum addresses.

PPR defines data ownership and stewardship between patient and provider, access information such as data pointers and access permissions, and mining bounties for researchers. Finally, SC is a ledger with all references to PPR contracts. Each node consists of backend API library, Ethereum client, database gatekeeper, and EHR manager and is managed via a web user interface. The backend library aids communication between the user interface and the Ethereum client. The database gatekeeper verifies a query request and returns the query result to the user. When a provider adds a new patient, the backend library communicates it to the Ethereum client. The Ethereum client utilizes RC to create a new PPR contract defining the provider's data stewardship for the patient's medical record. The patient's SC gets updated with the new PPR contract. Hence, every time a provider requests stewardship for their medical data, the patient will get an alert to grant or deny the request. This way, patients keep ownership and complete control of their records.

S. Abdullah et el. [9] suggested potential blockchain use cases in radiology such as authentication and verification, administrative and governance, and research and machine learning. Examples of blockchain usage for authentication and verification include image sharing access control, health data network authentication, and imaging data verification. Blockchain can also manage claim adjudication, billing, supply chain, imaging equipment maintenance, and inspection record-keeping for administrative and governance purposes. In addition, blockchain can enhance data sharing for research, clinical trials, machine learning training, and AI model execution. An imaging sharing company, NucleusHealth (San Diego, CA, USA) [10] uses Ethereum blockchain to create access control rules for image sharing between two systems. Patient health information (PHI) is stored on the blockchain, nor is medical imaging data transferred within the network. Instead, this Ethereum-based image sharing application allows image viewing utilizing DICOM Web URL and DICOMWeb RESTful API while keeping the images at the site's resource location. In addition, unique exam IDs are stored in the blockchain with the patient attached to the exam, and smart contract allows patients to share their images by granting access.

MJH Faruk et al. [11], [12] introduced Electronic Data Management (EHR) by adopting both Ethereum and Hyperledger Fabric in two different studies. Presented blockchain-based data storing and sharing platform offer patients, healthcare providers and authorized third parties to store, share and access the EHR data seamlessly. The framework ensures security through smart credential management, transparency by using smart contracts, and finally incorruptible nature of the medical record storage network. M. Jabarulla and H-N Lee [13] proposed a patient-centric image management system using an Ethereum blockchain and Inter-Planetary File System (IPFS)-based decentralized framework to solve the current system's challenges, including privacy, security, access flexibility, and costs. It stores and shares access to medical images within an open distributed network while offering transparency and patients' complete control over their images. In addition, smart contract was used to enable patient-centric access management, through which patients can grant/revoke access to their providers.

B. Shen et el. [14] mentioned scalability and latency as one of the top barriers to adopting existing blockchain applications and suggested a session-based data sharing using blockchain to address the problem. MedChain connects healthcare organizations through a decentralized network and enables session-based data sharing among peers. The network has two different peer nodes: super peers and edge peers. Super peers provide computing and storage resources for the data-sharing infrastructure, while edge peers store patient data. Modules of super-peer consist of blockchain service, directory service, and healthcare database. Blockchain server keeps records of data sharing to provide data integrity and auditing while the directory server manages patient data inventory and data sharing sessions, and these servers form sub-networks. Healthcare database stores patient data. The event is recorded in the blockchain when new data is generated, and data description is added to the directory service. When providers request a patient's data, they select the data from their inventory to share, creating a session. The blockchain service adds the session in a new block and issues a session ID with which Providers can access the data.

Healthcare blockchain applications mentioned above emphasize patient-centric healthcare data sharing and management. While there have been use cases and research on improving patient-centricity using blockchain, few researchers have taken utilizing blockchain to improve workflow efficiency into consideration. In this paper, we want to concentrate on improving workflow efficiency in teleradiology with blockchain technology. Our proposed blockchain-based medical image sharing and automated critical-results notification application will address the current challenges of teleradiology workflow and provide a potential solution for improved efficiency in the workflow.

## B. Overview of Hyperledger Fabric

Hyperledger Fabric is an open-source framework for permissioned blockchain networks [15], [16]. A permissioned network operates on a closed network and all participants have known identities, contrary to a public network where anyone can enter the network to see the transaction records [17]. Furthermore, Hyperledger Fabric offers technical advantages to achieve enhanced performance, scalability, security, and privacy for healthcare data. For example, its modular architecture provides optimized performance, scalability, and levels of trust, and data partitioning through channels provides enhanced confidentiality and privacy of data [18]. Hyperledger network consists of peer nodes, certificate authorities (CA), membership service providers (MSP), chaincode (smart contract), channels, shared ledger: world state and transaction log, assets, organizations, and ordering service [19].

- Immutability and integrity of data: Each member has a copy of the ledger. Distributing ledger among members prevents a single point of failure and guarantees immutability and integrity.
- Identity management: MSP manages user authentication and authorization through digital certificates.
- Privacy and confidentiality: Channels can be created among a group of network members to keep specific transaction data private and confidential.
- Speed and accuracy: Smart contract define business logic as executable code and automates transactions eliminating the need for intermediaries.

## III. CHALLENGES IN CURRENT TELERADIOLOGY WORKFLOW

Radiological exams are one of the key diagnostic tools as they provide anatomic details by visualizing the inner body structures. Therefore, getting a radiology report on time is crucial for timely patient care and disease management. Reducing radiology report turnaround time can significantly impact patient outcomes, especially in critical conditions like stroke, acute pulmonary embolism, or intracranial hemorrhage. We analyzed teleradiology's current workflow to identify pain points that cause delays in radiology report turnaround time and interruptions in radiologists' workflow. The workflow analysis was performed based on the author's direct work experience as a radiology worklist manager and interviews with a radiologist and a support team member.

### A. Challenges in current workflow

Image unavailability and delayed critical-results communication are among the most substantial causes of prolonged turnaround time in the current teleradiology workflow.

- Image availability: As mentioned in the Introduction, teleradiologists do not have direct access to the sending site's RIS or PACS [3]. As a result, they can only review images once the originating site sends them, which creates potential inefficiencies in the workflow regarding image availability. First, radiologists must communicate the need for additional data, including missing images, additional image processing, and prior reports. Second, it is hard for offsite radiologists to surmise the wholeness of data when a study comes over with fewer images than standard image sets for the study because there is no way to check if the study contains fewer images due to data lost in transit or the sending site's imaging protocol. Radiologists must request to check data completeness as well.

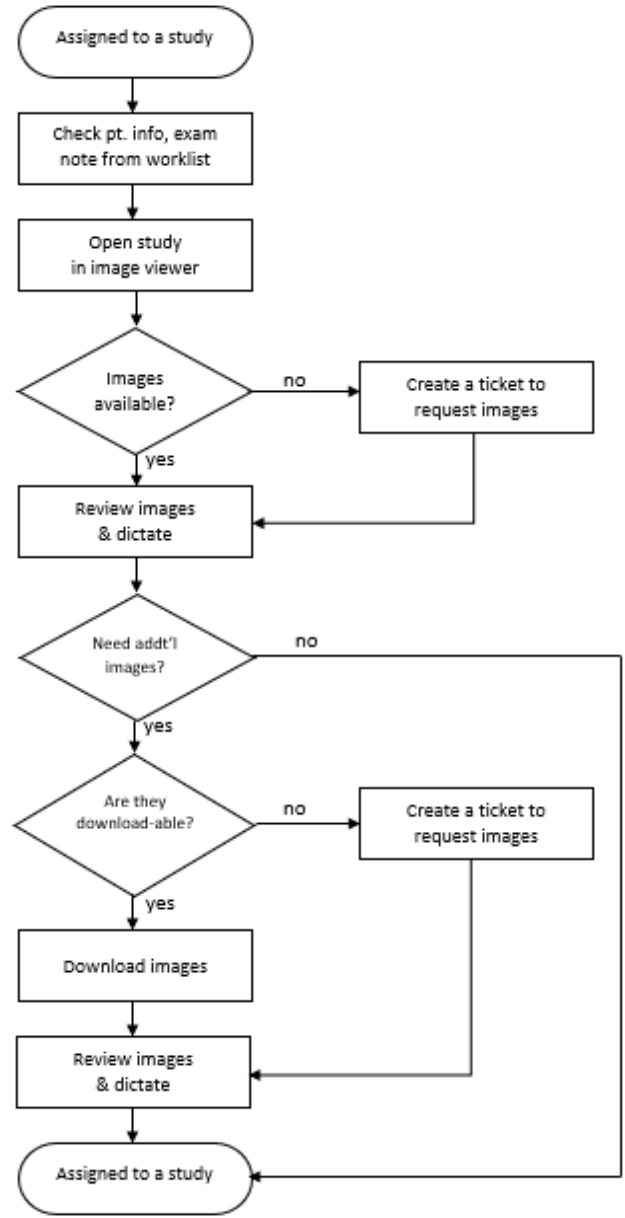

Figure 1: Current workflow for image request ticket

As shown in Fig.1, once a teleradiologist requests missing/additional images by creating a ticket from the worklist, a support team member contacts the site asking to resend the images. Upon receiving the requested images, the support team member closes the ticket and makes the exam available for the radiologist to interpret. This process may sound simple; however, it can easily take up to 20-30 minutes in real-world scenarios, involving multiple personnel requiring phone calls and sometimes voice mails. Moreover, getting additional phone calls and resending images are an extra workload for clinicians and can negatively impact their workflow.

- Critical-results communication: Critical conditions require immediate medical attention. Therefore, these critical findings need to be conveyed promptly. The current process for critical result communication is the same as requesting images mentioned above; a teleradiologist requests a critical conference call through a ticket from the worklist. Then, a support team member contacts the site to connect the teleradiologist with the provider. Although critical conference call request prioritizes others, delays still happen due to difficulty reaching the referring physician or simply from a large ticket volume.

Blockchain technology can be a potential solution to address these problems. Our proposed blockchain-based medical image sharing and automated critical-results notification model offers a fast and efficient way to share medical images and communicate critical findings to reduce radiology report turnaround time and improve radiologists' workflow efficiency.

## IV. PROPOSED ARCHITECTURE

We have developed an architecture to improve image availability and automate critical-results notification. The proposed architecture provides a private and permissioned network with all participants having known identities and its technical advantages support enhanced performance, scalability, security, and privacy. Hyperledger Fabric can be a suitable implementation for the proposed architecture.

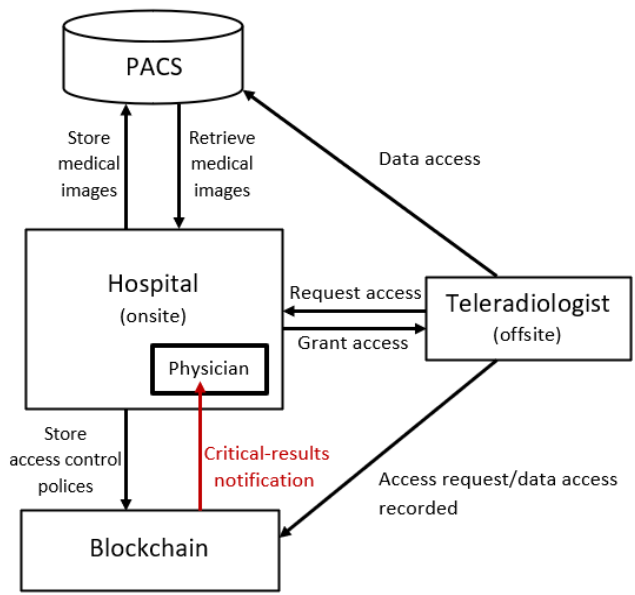

Figure 2: Proposed architecture

Figure 2. shows our conceptual model for blockchain-based medical image sharing and automated critical-results notification application. This model consists of users (radiologists, physicians), organizations (hospitals), onsite PACS, and blockchain networks. All medical images would be stored off-blockchain in the onsite Picture Archiving and Communication System (PACS) and blockchain controls and manages access to medical images stored in the onsite PACS [20], [21]. Registered radiologists can request access to medical images. Peer nodes in the hospital verify the authenticity of the access request. If access is granted, the radiologist will be provided with a link to view the requested images. Hash is created for every request and data access and gets recorded in a block. The blockchain is a chronological log of blocks that are interlinked by block hash. Each block consists of a block header and block data. The Block header contains hashes of the current block and previous block, and block data contains a list of transactions in chronological order as shown in Figure 3. We further discuss some features of our proposed architecture below.

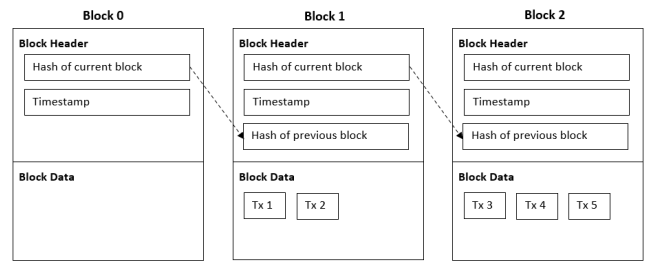

Figure 3: Structure of blockchain

### A. Registration and enrollment

All users must register to the blockchain network and enroll in an organization. Fabric certificate authority (CA) admin proves the user's identity using public-private key pair and issues a user ID. Membership service providers (MSPs) link the user's ID to the membership of an organization and define the user's role and permissions inside the organization.

### B. Data access

Medical images are stored in the site's PACS, and exam IDs are stored in blockchain. Offsite radiologists can request access through his/her user application and a link to view the images will be provided once the request is verified by peers in the blockchain network. Records of access requests and data access are added to a block in the form of hash and the ledger gets updated with a new block.

### C. Automated critical-results notification

Smart contracts are invoked to send an alert to the referring physician when pre-arranged keywords are detected in the radiology report.

## V. DISCUSSION

We demonstrated the current teleradiology workflow in section III. A radiologist works with three applications: worklist, image viewer, and dictation application. Radiologists can check patient history and exam information, get a link to view the images, and open a ticket to communicate needs for images or critical conference calls from the worklist. Support team members manage the worklist and contact the site to get the requested images or connect the radiologist to the referring physician based on the request on a ticket.

We believe that our blockchain-based medical image sharing and automated critical-results notification application can enhance efficiency in workflow by eliminating the need for intermediaries. With our application, radiologists will request images using a user application and be provided with a link to view the images after the authentication and validation process. A comparison of the two systems is summarized in Figure 4.

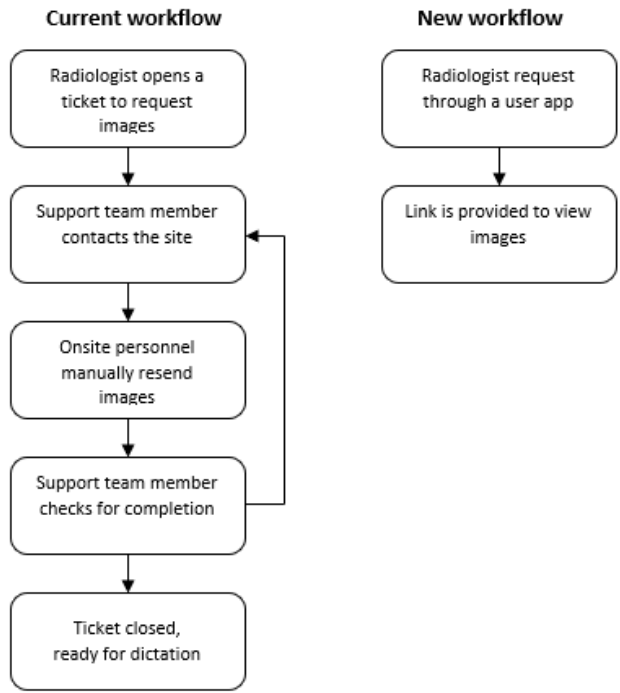

Figure 4: Workflow comparison

One of the limitations of our proposed architecture is that we did not include patients as an entity as we developed the model emphasizing workflow efficiency. However, patients can be added to the network to view their medical images and request to share images with their physicians. This will benefit patients by eliminating the need for storing medical images in hard copies (e.g., CD or DVD). There are also limitations to implementing the new application, including organizational and culture change, initial expenses, and regulations. Although the modular architecture of our proposed model and usage of the originating site's PACS infrastructure facilitates the implementation and aids regulatory compliance, further research on governance and HIPAA compliance is required to optimize the adoption of the new application [22], [23].

## VI. CONCLUSION

In this paper, we proposed a blockchain-based medical image sharing and automated critical-results notification to improve efficiency in teleradiology workflow and reduce radiology report turnaround time for quality patient care. We presented a detailed architectural framework. We also discussed challenges in the current workflow: prolonged turnaround time caused by image unavailability and delayed critical result communication and how blockchain technology can be used to address these issues. As part of future work, we aim to implement our proposed model and measure performance metrics, including efficiency, and average response time to calculate how much time can be saved by utilizing our model.